# Thermoelectric properties of a nanocontact

*Keivan Esfarjani\*[1], Mona Zebarjadi[1], Ali Shakouri[2] and Yoshiyuki Kawazoe[3]*

[1]Department of Physics, Sharif University of Technology, Tehran 11365-9161, Iran

[2]Department of Electrical Engineering, UCSC, Santa Cruz, 95064 USA

[3]IMR, Tohoku University, Sendai, 980-8577 Japan

(\*) k1@sharif.edu

ABSTRACT

Thermoelectric properties of a nanocontact made of two capped single wall nanotubes (SWNT) are calculated within the tight-binding approximation and Green's function method. It is found that semiconducting nanotubes can have high Seebeck coefficient very near the actual Fermi energy. This in turn leads to very high figures of merit easily exceeding one. Modifying the properties by shifting the chemical potential can be achieved by doping or application of a (nano-)gate voltage. The presence of impurities near the contact also strongly modify the properties and this can be taken advantage of, in order to design devices with high thermopower and figures of merit.

KEYWORDS

MANUSCRIPT TEXT

Today's electronic devices are becoming increasingly small and dense. Extraction of the produced heat from them is therefore of paramount importance. Materials with high thermoelectric properties, namely the figure of merit, can be used for this purpose[1]. It has been known[2] that reducing the dimensionality can increase the thermopower. Indeed, in lower dimensions, singular features appear in the density of states (DOS), and this can lead to large variations of the Fermi-energy-dependent conductivity. The Seebeck coefficient or thermopower, being proportional to the logarithmic derivative of the latter, can therefore become large, and thereby lead to high figures of merit, $ZT$. As the DOS of metallic wires is smooth near the Fermi level, metals can not have a large Seebeck coefficient. Semiconductors, for which the variation of the DOS near the Fermi level is non analytic, are therefore much better candidates for achieving high $ZT$. The problem with semiconductors being that they can have high thermal conductivity (due to phonons) which can reduce $ZT$ since the former is defined as $ZT = \sigma S^2 T / \kappa$ where $\sigma$, $S$ and $\kappa$ are respectively the electrical conductivity, the Seebeck coefficient, and the thermal conductivity. Search for low dimensional semiconducting materials with high figures of merit has thereby been undertaken since 1993[1,2].

Sun et al.[3] calculated the thermoelectric properties of Bi nanowires within the semiclassical theory, using the relaxation time approximation, anisotropic effective masses and non-parabolicity of the energy dispersion, and found that for appropriate doping, the figure of merit of a 10 nm wide nanowire can reach 1.5 and increases as the wire width is decreased. On the experimental side, Hsu et al.[4] have synthesized alloys containing nanometer size metallic grains embedded in a semiconducting matrix, reaching a $ZT$ of the order of 2.2, implying again the important role of quantum confinement. A $ZT$ value of 2.4 was also observed in $Bi_2Te_3$ and $Sb_2Te_3$ superlattices[5]. On the other hand, Lyeo et al.[6], have used the high thermopower of a nanocontact to map out the thermopower profile of a semiconductor substrate by their Scanning Thermo Electric Microscopy (SThEM) device. It becomes clear then that the quantum confinement present in nanoscale systems can yield interesting thermoelectric properties, which can be used in order to design cooling devices.

In this work, to illustrate further the important thermoelectric properties of nanocontacts, we have considered nanocontacts formed of different configurations of two capped single wall carbon nanotubes (SWCNT) of metallic and also semiconducting nature, and have calculated their conductance, thermopower and figure of merit. Some typical configurations of two (10,0) semiconducting SNCNT used in our calculations, before and after doing molecular dynamics (MD) can be seen in Fig.1.

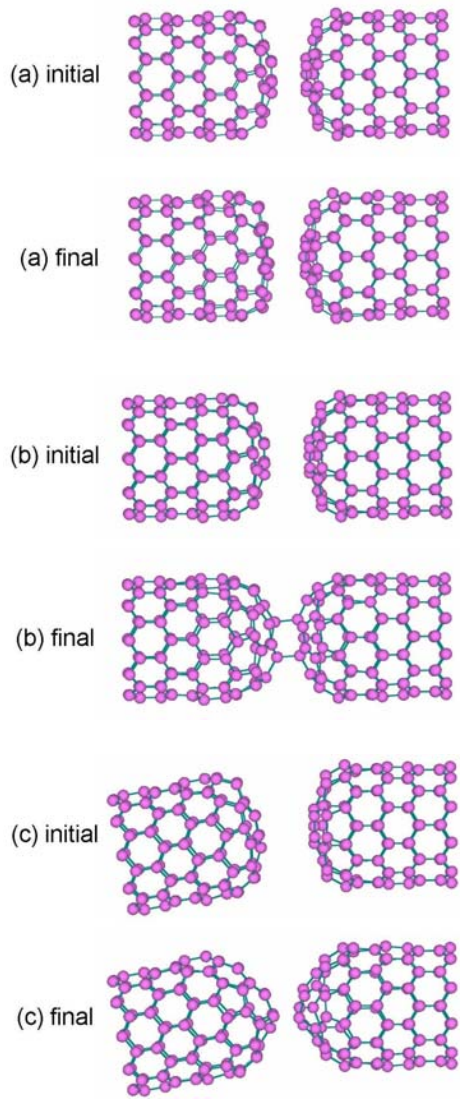

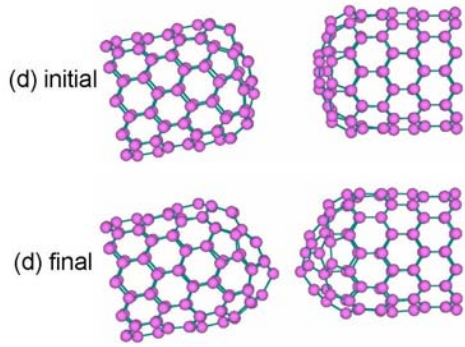

Figure 1. The initial and a typical MD snapshot (called here final) of two semiconducting (10,0) capped nanotubes forming a nanocontact. In order not to take any symmetric configuration, one of the tubes has been rotated and tilted with respect to the other.

To include the effect of vibrations, we have used an adiabatic approximation by taking a configuration average of the transmission coefficient over many (50) uncorrelated snapshots obtained from a molecular dynamics (MD) simulation. The resulting thermoelectric properties were obtained by configuration-averaging the properties calculated from each MD-generated snapshot. In order to calculate the linear transport properties of the nanocontact, the Landauer-Buttiker formula[7,8], in which the transmission coefficient is calculated by using the Green's function method[9,10,11] was used. The formalism is exact in the coherent transport limit where inelastic scattering and phase breaking mechanisms such as electron-phonon interaction are absent. The formulation is within a tight-binding approximation where, for simplicity, only one $\pi$ orbital was considered for each carbon atom. The effect of the other orbitals forming an sp2 hybridization ($\sigma$ bonds) becomes important away from the Fermi energy. The leads are assumed to be continuations of the perfect nanotubes over which it is assumed that there is no voltage or temperature drop. The two-probe Landauer formula can then be used for this purpose. One needs to take the "nanocontact" region long enough so that all potential and temperature drops occur within it. Due to memory and CPU time limitations, we have considered a central region containing 120 atoms. Using infinitesimal voltage and temperature gradient drops across the sample makes our approximation valid. Within the linear response limit, the thermoelectric properties of a nanosystem can be derived from the definitions of the electrical and thermal currents[7,8]:

$$I_e = \frac{2q}{h} \int dE\, T(E)\, [f_L(E) - f_R(E)]$$
$$I_Q = \frac{2}{h} \int dE\, T(E)\, [f_L(E) - f_R(E)](E - \mu) \quad (1)$$

where the factor of 2 is for the spin, q is the carrier charge, $T(E)$ is the transmission coefficient of the device, and $f_L$ and $f_R$ are respectively the distribution functions of the left and right reservoirs with chemical potentials $\mu_L$ and $\mu_R$. In this work, we are interested in the linear response of the system i.e. we assume that $\mu_L - \mu_R$ as well as $T_L - T_R$ are infinitesimally small quantities so that the currents are linear in $\mu_L - \mu_R$ and $T_L - T_R$. Therefore the chemical potential $\mu$ in the thermal current is defined as the average of the left and right chemical potentials. Furthermore the thermoelectric response functions can be obtained from the ground state properties of the system. The response of the system is thus defined by the following matrix:

$$\begin{pmatrix} I_e \\ I_Q \end{pmatrix} = \begin{pmatrix} q^2 K_0 & q K_1 \\ q K_1 & K_2 \end{pmatrix} \begin{pmatrix} -\Delta V \\ -\Delta T / T \end{pmatrix} \quad (2)$$

By using a Taylor expansion in powers of $\Delta V$ and $\Delta T$ in eqs. 1, it can easily be shown that the coefficients $K_n$ are given by the following integrals

$$K_n = \frac{2}{h} \int dE\, T(E) \left( -\frac{\partial f}{\partial E} \right) (E - \mu)^n \quad (3)$$

The derivation is trivial and is very similar to the well-known and standard semiclassical one used in textbooks[12], except that here the function $T(E)$ has replaced the energy-dependent Drude conductivity $\sigma(E) = 2 \sum_k v_k v_k \tau_k\, \delta(E - \varepsilon_k)$. To our knowledge, the above relations were first derived and discussed by Sivan and Imry[13]. The transport properties defined below, can then be related to the integrals $K_n$ which strongly depend on the analytic properties of the transmission coefficient near the Fermi level.

- Conductance : $G = -I_e / \Delta V \vert_{\Delta T = 0} = q^2 K_0$

- Peltier Coefficient : $\Pi = I_Q / I_e \vert_{\Delta T=0} = K_1 / qK_0$

- Thermopower : $S = -\Delta V / \Delta T \vert_{I_e=0} = K_1 / qT K_0$

- Thermal conductance : $\kappa = -I_Q / \Delta T \vert_{I_e=0} = (K_2 - K_1^2 / K_0)/T$

If the transmission function $T(E)$ is analytical, such as in metals, one can use its Taylor expansion around the chemical potential in order to obtain its temperature dependence by using the Sommerfeld expansion. Mott's formula, relating the thermopower to the logarithmic derivative of the conductivity in metals, as well as Wiedemann-Franz law, expressing the proportionality of the thermal to the electrical conductivity in metals can be proven by using such an expansion. Finally, the figure of merit is simply given by

$$ZT = G S^2 T / \kappa = K_1^2 /(K_2 K_0 - K_1^2) \quad (4)$$

Note that, although in the definition of $ZT$, the full thermal conductivity is introduced, in the second part of eq. 4, we have only put the electronic contribution to $\kappa$ and the phononic part is omitted. This is a good approximation in metals, but in semiconductors or insulators, the largest contribution in $\kappa$ is due to phonons. In the considered nanocontact (see Fig. 1), however, since the system is split into two, we do not expect that $\kappa$ due to phonons would have a significant contribution, even if the considered nanotube is semiconducting.

For the considered nanocontact, as mentioned, we generated uncorrelated configurations from 50 MD snapshots run at 300 K, for which all the above properties were calculated and then configuration-averaged. For the snapshot shown in Fig. 1, the initial cap-cap separation was chosen to be as large as 3.2 Å, but it is worth noticing that during the MD run, due to the weak Van der Waals interaction between the caps, the tubes are slightly elongated and form almost a chemical bond, so that the cap-cap distance in the 50 snapshots varies from 1.7 to 2.2 Å.

The calculations were done for both a pair of (5,5) metallic tubes and a two pairs of (8,0) and (10,0) semiconducting tubes all at room temperature (in the Fermi-Dirac function in eq. 3, $k_B T = 0.025\,eV$). Although results are similar, they differ in the fact that the interesting features such as large figure of merit (see Fig. 2), occur for chemical potentials near the actual Fermi level of the semiconducting tubes only. So, in what follows, we will focus on the results obtained for the (10,0) tubes.

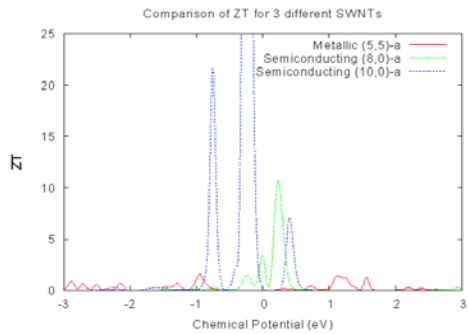

Figure 2. Figures of Merit for one (5,5) metallic, and two (8,0) and (10,0) semiconducting nanotube contacts, plotted as a function of the chemical potential. For the sake of clarity, the y axis was cut at 25, but $ZT$ of the (10,0) tube reaches the value of 100. Although $ZT$ for the metallic tubes is of the order of 1, that of semiconducting tubes can easily exceed 5, especially near the actual Fermi level of the tubes, which is zero.

The configurations were generated for two initial separations of the nanotubes denoted by (a) and (b). In structure (a), the initial separation is 2.4 Å, and in (b) it is increased to 3.2 Å (see Fig. 1).

After some relaxation time, however, the cap-cap distance is reduced to below 1.8 Å as can be seen in Fig.1. In other runs (structures c and d of Fig. 1), the tubes on each side of the junction were also slightly tilted and rotated so that no accidental symmetry affects the results.

In Fig. 3, conductances are compared for the (10,0) tubes placed at two different initial distances and two different orientations. (a) is clearly in the tunneling regime and (b) (c) and (d) have stronger coupling to leads as $G$ can be of the order of $G_0$ for values of the chemical potential away from the gap.

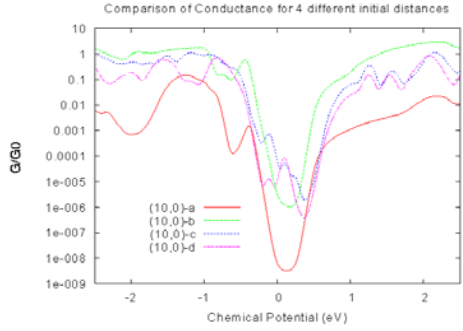

Figure 3. Log of the averaged conductance of the (10,0) tubes for the four separations (a),(b),(c), and (d), in units of $G_0 = 2e^2/h$ plotted as a function of the chemical potential. One can notice that the magnitude in case (a) is much reduced compared to (b) and others, meaning that the transport in (a) is tunneling dominated ( $G \approx 0.01 G_0$ ), whereas that in (b) is in the strong-coupling-to-leads regime( $G \approx G_0$ ).

The semiconducting gap near energy 0 is apparent, and one can notice very sharp changes in $G$ by 2 or 3 orders of magnitudes within 0.5 eV near the gap. Such large changes are responsible for large Seebeck coefficients, as the latter is the logarithmic derivative (Mott's formula) of the conductance. Although there is still a large variation in the four conductances near the gap, conductances of the four samples are quite different reflecting the mesoscopic character of the considered system.

Simply speaking, the nanocontact can be viewed as a barrier in the transport of electrons in a quantum wire. Interference effects due to multiple reflections at the contact walls, therefore, strongly affect the transport and create the observed fluctuations in the conductance.

The variations of the thermopower for structures (a) and (b) are displayed in Fig. 4.

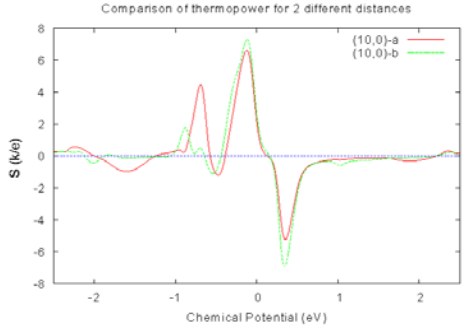

Figure 4. Seebeck coefficient for the two separations (a) and (b), in units of $k_B/e$ plotted as a function of the chemical potential. One can notice the large fluctuations in the thermopower (due to mesoscopic conductance fluctuations), and its sign changes every time the conductivity is increasing and decreasing. For the sake of clarity only two typical curves are shown.

Both tubes display large values for $S$, as large as 7 $k_B/e$. The actual chemical potential being equal to zero, however, only values of $S$ or $G$ or $\kappa$ near $\mu = 0$ are experimentally accessible. If hole-doped, the considered semiconducting tubes can achieve large thermopowers. As a comparison, standard semiconductors have $S \approx k_B/e$ and metals have $S \approx 0.01 k_B/e$. Therefore this leads to a figure of merit which is a factor of almost 50 times larger compared to usual semiconductors, all other quantities assumed to be equal. The other factor influencing the figure of merit is the Lorentz number: $\kappa/GT$, which according to the Wiedemann-Franz law is a constant for metals. In semiconductors, however, it can fluctuate as the Fig. 5 shows. The difference from 1 of the ratio $3\kappa/\pi^2 G k_B^2 T$ shows a deviation from Wiedemann-Franz (WF) law. As can be seen in Fig. 5, this number varies from 0.1 to 10, and therefore, we can see large deviations from the WF law in semiconductor nanotube contacts.

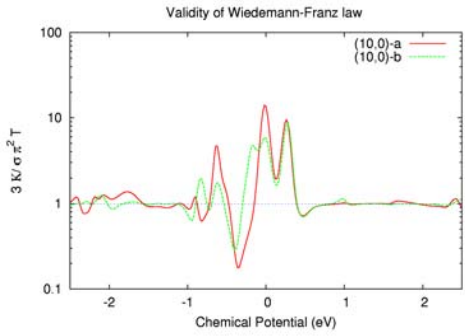

Figure 5. Logarithm of $3\kappa/\pi^2 G k_B^2 T$ versus the chemical potential. Legends are similar to the previous figures.

Finally, the figure of merit of the 4 configurations is plotted versus the chemical potential in Fig. 6. One can see that near $\mu = 0$ the structure (a) can have figures of merit as large as 100! This can be traced back to a large thermopower in this energy range, and also a deviation from the WF law.

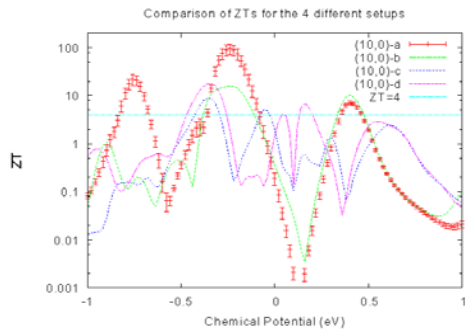

Figure 6. Figure of merit versus the chemical potential. Legends are similar to the previous figures. For all configurations, it can reach values exceeding 10 just below the natural Fermi energy of the tubes. For the structure (a) the error bars are also shown. $ZT = 4$ of conventional coolers is also shown for comparison.

The advantage of this semiconducting set up compared to other semiconducting thermoelectric devices is that the phonon contribution to the thermal conductivity, in the case where the separation is large, is zero and thus the denominator in $ZT$ does not need corrections and it remains small. It can also be seen that for all 4 configurations $ZT$ can easily exceed 4 if the tubes are hole-doped. On the figure, the error bars for structure (a) are the error in mean, and since the average is over 50 configurations, the actual fluctuations in *ZT* are almost 7 times larger than the error bars. One can conclude from this study, that for many arbitrarily chosen configurations the average figure of merit can easily exceed 10.

There are some simplifying assumptions and approximations in this model. Namely it is a one $\pi$ orbital per carbon calculation of the coherent conductance, neglecting the electron-phonon and electron-electron interactions. We believe, based on the present calculations, however, that large values for the Seebeck coefficient and also the figure of merit can be achieved in such systems, and as we showed these features are quite robust with respect to atomic vibrations and different orientations of the two tubes. The present calculation is a proof of the concept. In an actual device, several such setups might be needed to be put in series in order to achieve reasonable amount of cooling. The experimental realization of the proposed device might not be very easy, but we believe that other similar setups made of semiconducting quantum wire contacts will also display similar properties as the work by Hsu et al. on granular metals in a semiconductor matrix is already a witness to it. In their case, one is also in the tunneling regime, but presumably the contribution of phonons to the thermal conductivity brings down the figure of merit of the device. Finally, it should be added that the present results are obtained within the linear response approximation and assume infinitesimally small potential and temperature gradients applied to the device.

ACKNOWLEDGMENT

K.E. would like to acknowledge the Iranian Ministry of Science and Technology for financial support, and the Japanese Society for Promotion of Science (JSPS) for funding his trip to the Institute for Materials Research where part of this work was done. Dr. Amir. A. Farajian is also acknowledged for useful comments and a critical reading of the manuscript.

---

[1] Mahan, G.D. Jour. Appl. Phys.**1994**, 76, 4362; Mahan, G. D. Solid State Physics; Academic Press, New York, **1998**, 51, 81.